\title{CS224W project}
\documentclass[conference]{IEEEtran}

\ifCLASSINFOpdf
\else
\fi
\usepackage{graphicx}
\usepackage{epstopdf} 
\usepackage{booktabs}
\usepackage[noend]{algpseudocode}
\usepackage{verbatim}

\usepackage[cmex10]{amsmath}

\makeatletter
\def\BState{\State\hskip-\ALG@thistlm}
\makeatother

\begin{document}
%
\title{Structural Analysis of Criminal Network and Predicting Hidden Links using Machine Learning}

\author{\IEEEauthorblockN{Emrah Budur}
\IEEEauthorblockA{
Stanford University\\
Stanford, California 94305\\
Email: emrah@stanford.edu}
\and
\IEEEauthorblockN{Seungmin Lee}
\IEEEauthorblockA{School of Electrical Engineering\\
Stanford University\\
Stanford, California 94305\\
Email: smlee729@stanford.edu}
\and
\IEEEauthorblockN{Vein S Kong}
\IEEEauthorblockA{
Stanford University\\
Stanford, California 94305\\
Email: skong@umich.edu}}


\maketitle

\begin{abstract}
Analysis of criminal networks is inherently difficult because of the nature of the topic.  Criminal networks are covert and  most of the information is not publicly available.  This leads to small datasets available for analysis. The available criminal network datasets consists of entities, i.e. individual or organizations, which are linked to each other.  The links between entities indicates that there is a connection between these entities such as involvement in the same criminal event, having commercial ties, and/or memberships in the same criminal organization. Because of incognito criminal activities, there could be many hidden links from entities to entities, which makes the publicly available criminal networks incomplete.  Revealing hidden links introduces new information, e.g. affiliation of a suspected individual with a criminal organization, which may not be known with public information. \textit{What will we be able to find if we can run analysis on a larger dataset and use link prediction to reveal the implicit connections?}  We plan to answer this question by using a dataset that is an order of magnitude more than what is used in most criminal networks analysis.  And by using machine learning techniques, we will convert a link prediction problem to a binary classification problem. We plan to reveal hidden links and potentially hidden key attributes of the criminal network.  With a more complete picture of the network, we can potentially use this data to thwart criminal organizations and/or take a Pareto approach in targeting key nodes.  We conclude our analysis with an effective destruction strategy to weaken criminal networks and prove the effectiveness of revealing hidden links when attacking to criminal networks.

\end{abstract}

\IEEEpeerreviewmaketitle

\section{Introduction}

This paper will be focusing on predicting hidden links in the context of criminal networks.  Because of the nature of the subject: data and links for each node might be hard to get, or might in fact be hidden.  For our project we surveyed a few papers: three of them deal with link prediction in the context of social networks, and two of them focusing on how social network analysis can be used to help understand crime and terror organizations.

\subsection{Overview}

The six papers we surveyed will give us a background on how link prediction is done for social networks and will give us a better idea on how to model and use network analysis to potentially find hidden links in crime networks.  

In the Clauset et al. paper, it is focused on hierarchical structure and link prediction based on the hierarchical structure \cite{Clauset2008}.  The paper selected three networks for analysis and using hierarchical random graph they were able to have similar average degree, clustering coefficient, and vertex-vertex as the selected networks.  Most criminal networks will likely fit into the hierarchical category.  It generates many hierarchical networks and use those randomly generated models to find high probable edges.  It also has a comparison of area under curve (AUC) of the different link prediction methods which is really valuable.  

The Hasan and Zaki paper focuses on surveying different link prediction models used in social networks  \cite{Hasan2011}.  They categorized them in four different categories: 1) Featured-based 2) Bayesian 3) Probabilistic 4) Linear Algebraic.  They talk about the approaches in high level and does a brief review of all of them.  They have many good feature suggestions, such as common neighbors, Jaccard coefficient, Katz, and shortest path distance. 

Liben-Nowell and Kleinberg, takes Hasan and Zaki to the next level, by actually using different link prediction algorithms on actual data  \cite{Liben2003}\cite{Hasan2011}.  They selected research papers as the data and divide each into subject categories and ran a litany of link prediction algorithms on the dataset.  They noted that some algorithms jump out as performing well, but that there were a lot of room for improvement.  It also introduces us to a \textit{score(x,y)} feature which helps us determine if there's a link by \textit{embeddedness} or number of common neighbors.  We more than likely will take this as one of the approaches to finding hidden links.

Medina’s paper deals with using social network analysis (SNA) on terrorist networks, specifically al-Qaeda \cite{Medina2014}.  It uses basic SNA concepts.  It uses small-world and scale-free network to model al-Qaeda.  It tries to figure out whether al-Qaeda is hierarchical or decentralized.  It also removes two key nodes in the network to see how robust the network is.  Similarly in the Sparrow paper, it lays out some motivational points for using SNA in criminal networks \cite{Sparrow1991}.

In Backstrom and Leskovec, it talks about predicting links in a social network context, in this case, Facebook. The paper uses a \textit{supervised random walk} approach to predict missing links.  As mentined in the paper, research in security has recently recognized the role of social network analysis for this domain (e.g.,terrorist networks). In this context link prediction can be used to suggest the most likely links that may form in the future. Similarly, link prediction can also be used for prediction of missing or unobserved links in networks or to suggest which individuals may be working together even though their interaction has yet been directly observed. \cite{Backstrom_2011}

These papers are very pertinent to the course.  The topics of modeling using small-world or other models is prevalent in these papers and in our CS224W course.  Also with link prediction and survey of link prediction algorithms, this will be covered as a future topic in class.

The common themes throughout these papers are link prediction and trying out many different link prediction algorithms.  Also the attempt to model the network and getting attribute data on the network for future analysis is also prevalent throughout the paper.

\subsection{Critique}

Though the papers we reviewed gave us a good background on creating hidden links for criminal networks, it is not without fault.  There are a few missing gaps in the papers.  Most of them deal with the dataset selection and some gaps in the analysis.

Sparrow's work doesn't provide any mathematical approach or analysis \cite{Sparrow1991}.  It only suggests some concepts. 

Medina's work although pertinent to our research lacks a large enough dataset to provide any substantial findings  \cite{Medina2014}. The dataset of 381 nodes and 690 undirected links is a small sample size.  Since it is a public dataset of covert individuals, the dataset is incomplete.  The analysis on density seems to be flawed since the data is sparse.  It incorrectly uses the concept of small-world with just random graphs.  We think it should be more intentional in creating it with less randomness.  The analysis done on the network is pretty basic.  We would like to see more advanced analysis done on the dataset for completeness.

Liben-Nowell and Kleinberg's paper uses a diverse set of numerical analysis but doesn't have much explanation of the data or analysis \cite{Liben2003}. Also the dataset that they use is quite confined, the research paper's network would be different than a criminal network.  The scores that are calculated in the paper are independent, we think we can combine them for better analysis, Which we plan to do in our research.  The evaluation is performed over random networks, but we think it should use the test set error rate to evaluate it.

Clauset et al. focuses on hierarchical networks \cite{Clauset2008}.  Though most criminal networks are hierarchical, some may be decentralized, which would make Clauset's paper not as relevant.  The methods of common neighbors, shortest path length and product of degrees are correct in assortative networks, but can be misleading if our network is not assortative. The paper talks about $p_r$,which is probability  that node $r$ that a pair of right and left subtree of that node is connected ,but fails to give any real details of what $p_r$ should be for hierarchical vs non-hierarchical.

Hasan and Zaki's survey paper does just an overview of the methods without analyzing any real dataset \cite{Hasan2011}.

\section{Project proposal}

\subsection{Problem}
The common problem of the criminal network datasets is that they are incomplete and inherently covert by nature since public information doesn't hold complete information about the actors in the criminal events \cite{Sparrow1991}.  So, an existing direct relationship in the real world may not have been discovered yet, leading to hidden links in the network.  The hidden links may evolve into critical actions of future events or they may play a key role to extending existing networks to operate in a broader range. Moreover, they may result in revealing critical links and help to capture critical people that were wanted for a long time, like Saddam Hussein \cite{farnaz2003}. Therefore, sophisticated techniques are required to reveal the hidden links in criminal networks with high precision to solve an array of mission critical problems. 

\subsection{Dataset}

Most of the studies in criminal networks have used small sizes of datasets due to the scarcity of the accessible data in this field \cite{Sparrow1991}. Throughout this study we will use an anonymized large dataset that consists of 1.5 millions nodes worldwide with 4 millions undirected edges.  The nodes will include the publicly available sanction lists such as the famous US sanction list, namely OFAC \cite{ofac}.  Compared to datasets in the reviewed articles, our dataset has an order of magnitude more data, which will allow us to extract the real big picture of criminal networks worldwide.

\subsection{Scope Definition}
Our project consists of three main subproblems.

\subsubsection {Link Prediction Model using GBM}

\begin{figure}
\centering
        \includegraphics[width=9cm]{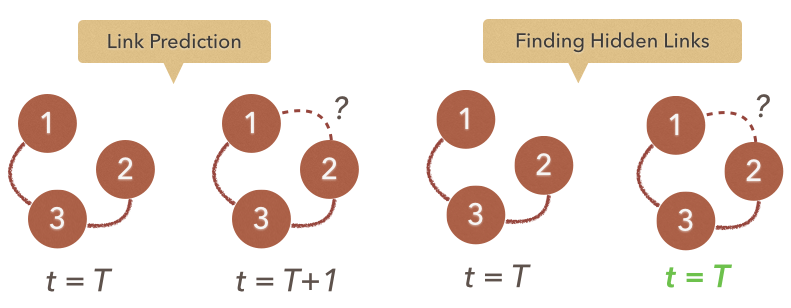}
    \caption{Comparison between Link Prediction And Finding Hidden Links}
    \label{fig:comparison_problem}
\end{figure}

We will apply a machine learning technique called Gradient Boosting Machine(GBM) on this problem. Unlike previous work for link prediction, we do not have transitional network data(network images at time $t$ and $t'$), we will make a supervised link prediction model based on the current network image. Fig \ref{fig:comparison_problem} depicts the concept. 

In order to apply GBM, we regarded this problem as a binary classification problem. We transformed a network image to a feature matrix with various metrics that are famously used in current link prediction studies.

We will analyze our model from various viewpoints and use this model as a base model for the future experiments.

\subsubsection {Finding Hidden Links in Criminal Network}
We will create corrupted networks, which will have 5\% to 50\% of edges removed from the original network, and then regard removed edges as hidden links. Now, we will test using our model to see how well our model trained in corrupted networks can find hidden links. This experiment is critical to find out whether we can find hidden links and the result would be important since we can detect hidden criminal connections or activities. 

We will also analyze various aspects of this experiment.

\subsubsection {Destroying Criminal Network}
Finally, we will go on our study with advanced analysis of our criminal network using our model. Probability of being an edge can be obtained by prediction using our model. We will use that probability as a weight on a specific edge, and apply pagerank algorithm on weighted network to get scores for each node. 
Then, we will analyse what method can destroy the criminal network as fast as possible as a result of this advanced analysis.  We will finally show that Weighted PageRank scores will help destruct the criminal network much smarter and much faster.  We will conclude that Weighted PageRank score can also be used as a  \textit{suspiciousness index} to use to target \textit{the vital few} of the network.

\subsection{Background}
This study requires a background on machine learning and social network analysis.  Below are a brief explanation of the essential subjects that are meant to provide a firm foundation for the method section.

\subsubsection{Terminology}

The word \textit{link} refers to an \textit{edge} between two nodes in the network.  The term $E$ refers to the set of all \textit{possible} edges regardless of the existence of the edge in the original network.

If there is an edge between two nodes in the original network, we call this edge a \textit{positive edge}.  The term $E^{+}$ refers to the set of all positive edges.

On the other hand if there is no edge between a particular pair of nodes in the original network, we use the term \textit{negative edge} to refer to the absence of the edge. The term $E^{-}$ refers to the set of all negative edges.  The term $E^{-}_d$ denotes the set of randomly chosen negative edges where each edge has shortest path length of $d$.

We used the term \textit{hidden edge} or \textit{hidden link} for a particular negative edge that we must \textit{predict} it as a positive edge.

\subsubsection{Link Prediction Metrics}
Table \ref{table:linkpredictionmetrics} shows a number of well-known metrics that are commonly used in link prediction between two nodes.  The formula $\varphi(i)$ represents the set of nodes that are neighbors of the node $i$ and $k_i$ represents the degree of the node $i$.

\begin{table}[h]
\centering
\begin{tabular}{@{}ll@{}}
\toprule
Metric                        & Definition \\ \midrule
Common Neighbors              & $S_{xy}=|\varphi(x)\cap \varphi(y)|$           \\
Salton Index                  & $S_{xy}=\dfrac{|\varphi(x)\cap \varphi(y)|}{\sqrt{k_x\times k_y}} $           \\
Leicht-Holme-Newman Index     & $S_{xy}=\dfrac{|\varphi(x)\cap \varphi(y)|}{k_x\times k_y} $           \\
Sorensen Index                & $S_{xy}=\dfrac{|\varphi(x)\cap \varphi(y)|}{k_x + k_y} $           \\
Jaccard Index                 & $S_{xy}=\dfrac{|\varphi(x)\cap \varphi(y)|}{|\varphi(x)\cup\varphi(y) |} $           \\
Hub Index                     & $S_{xy}=\dfrac{|\varphi(x)\cap \varphi(y)|}{\min(k_x, k_y)} $           \\
Preferential Attachment Index & $S_{xy}=k_x\times k_y$           \\
Adamic-Adar Index             & $\displaystyle S_{xy}=\sum_{z\in \varphi(x)\cap \varphi(y)}\dfrac{1}{\log{k_z}} $           \\ \bottomrule
\end{tabular}
\vspace{0.2cm}
\caption{Link Prediction Metrics}
\label{table:linkpredictionmetrics}
\end{table}

\subsubsection{Gradient Boosting Machine(GBM)} Gradient Boosting Machine is one of the most popular tree-based machine learning techniques. It is based on decision trees and it is used for both regression and classification problems.  GBM applies boosting methods for decision trees to reduce variance. Hence, the output of GBM has low variance.  

Compared to bagging method in other Tree-based methods, which creates many bootstrapped trees all at once and aggregating them, GBM keeps a long learner tree and grow it sequentially to avoid overfitting and high variance.  Instead of growing a single large decision tree to the data which results in potentially overfitting, the boosting approach learns slowly by iteratively correcting the error of the previous iteration. 

\begin{figure}
\centering
        \includegraphics[width=9cm]{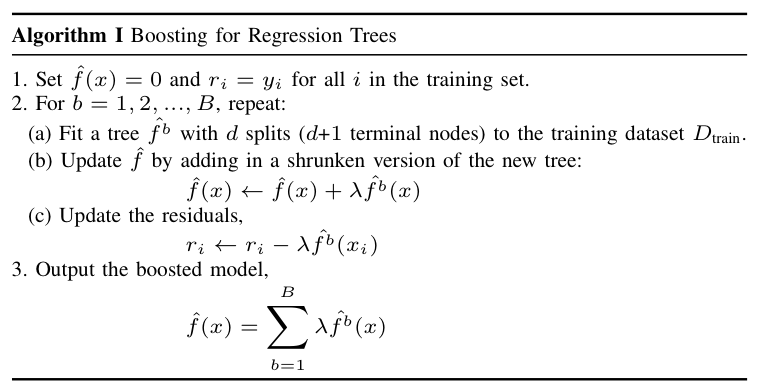}
    \caption{Algorithm : Boosting for Regression Trees \cite{gbmbook}.}
    \label{fig:gbm}
\end{figure}

As it is seen in algorithm in Fig \ref{fig:gbm}, GBM function has three tuning parameters:
\begin{itemize}
\item $B$: Is the number of trees. If we choose $B$ too large, boosting can overfit the data. So, a proper value of $B$ is chosen by validating the output of the function against the training dataset, which is known as \textit{cross validation}.
\item $\lambda$: Is the learning rate of the function. It is a small positive number which controls the rate at which boosting learns. There is a correlation between $\lambda$ and $B$. Small $\lambda$ requires  large $B$ in order to achieve good performance.
\item $d$: Is the number of splits in each tree, which controls the complexity of the tree structure. It is also known as \textit{the interaction depth}.  The bigger $d$ results in more interaction in each decision tree.
\end{itemize}

We used a set of parameters for GBM in Table \ref{table:parametersingbm} :
\begin{table}[h]
\centering
\begin{tabular}{@{}ll@{}}
\toprule
Parameters                        & Value \\ \midrule
Maximum Tree Depth              & 5           \\
Number of Trees                  & 50           \\
Fewest allowed observations in a leaf     & 10           \\
Number of Bins in a histogram                & 20           \\
Learning Rate                 & 0.1           \\
\bottomrule
\end{tabular}
\vspace{0.2cm}
\caption{Parameters used in GBM Model}
\label{table:parametersingbm}
\end{table}

\subsubsection{Base Data for Link Prediction Model}

In order to build a link prediction model, we designed a training and test data set. 

\begin{figure}
\centering
        \includegraphics[width=9cm]{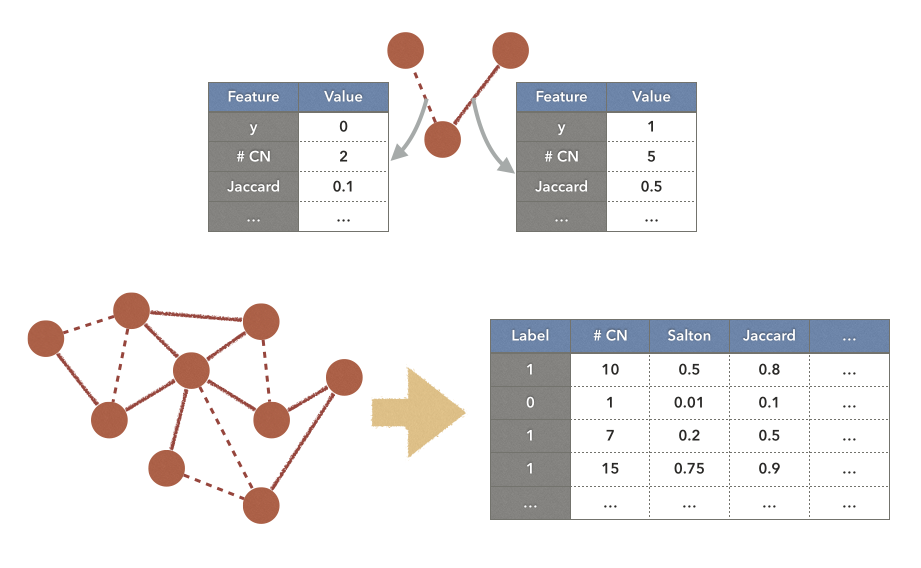}
    \caption{How to convert Network into Feature Matrix}
    \label{fig:converting_network}
\end{figure} 

First of all, we converted the criminal network into a feature matrix to apply machine learning. The concept is described in Fig \ref{fig:converting_network}.

We denoted the set of positive edges as $E^{+}$, and the set of nodes as $N$.  We denote the size of  nodes and positive edges as $|N|$ and $|E^{+}|$ respectively. The number of all possible edges is ${\left\vert N\right\vert \choose 2}$. It can be noted that ${\left\vert N\right\vert \choose 2} \gg |E^{+}|$. Consequently, the number of negative edges are much greater than the number of positive edges. Hence we chose a total of $|E^{+}|$ negative edges by randomly undersampling from $E^{-}$. In this way, we avoided bias in our machine learning model. Let's denote this dataset as $D$($D$ has total $2|E^+|$ edges; $|E^+|$ positive edges and $|E^+|$ negative edges). 

For the training phase, we prepared $75\%$ of $D$ to be used for training our model and we will refer it as $D_{\text{train}}$. We prepared the remaining $25\%$ of $D$ to be used for testing purpose and we will refer to it as $D_{\text{test}}$ for convenience. Of course, each of $D_{\text{train}}$ and $D_{\text{test}}$ has equal number of positive and negative edges.
 
After training GBM model using $D_{\text{train}}$, we will evaluate our model using the test set $D_{\text{test}}$ and calculate the resulting AUC.

\subsubsection{Undersampling}
Undersampling is a technique used in data analysis to adjust the class distribution of a specific data set. Undersampling randomly chooses some fraction of data from dominant classes to make equal numbers of class samples. For example, in our project, negative edges are overwhelmingly dominant, leading \textit{class imbalance}, so we undersampled those negative edges to make their numbers the same as the number of the positive edges. If we did not undersample, the model would be biased seriously with a tendency of predicting towards negative. 

We did an experiment to test whether our sampling method is accurate. First, since our dataset has so many nodes, considering all pairs is intractable, so we created a random graph of $10,000$ nodes following the power law with the same exponent that our criminal network has, $\alpha=3.6346$.
Now we compared the two model: one with undersampling negative edges to make 50:50=pos:neg, and the other without undersampling where the number of negative edges extremely dominates the number of positive edges.

\begin{figure}
\centering
        \includegraphics[width=7cm]{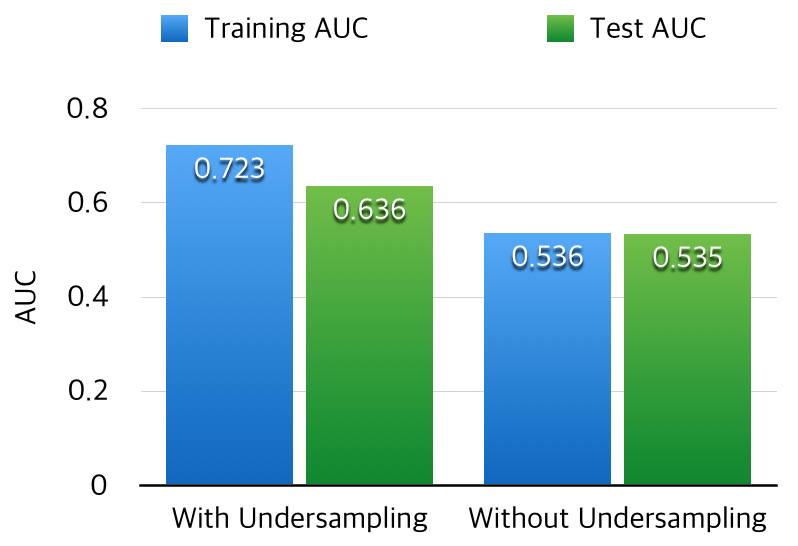}
    \caption{Comparing Model's Training and Test AUC between With Undersampling and Without Undersampling}
    \label{fig:auc_undersampling}
\end{figure} 

\begin{figure}
\centering
        \includegraphics[width=7cm]{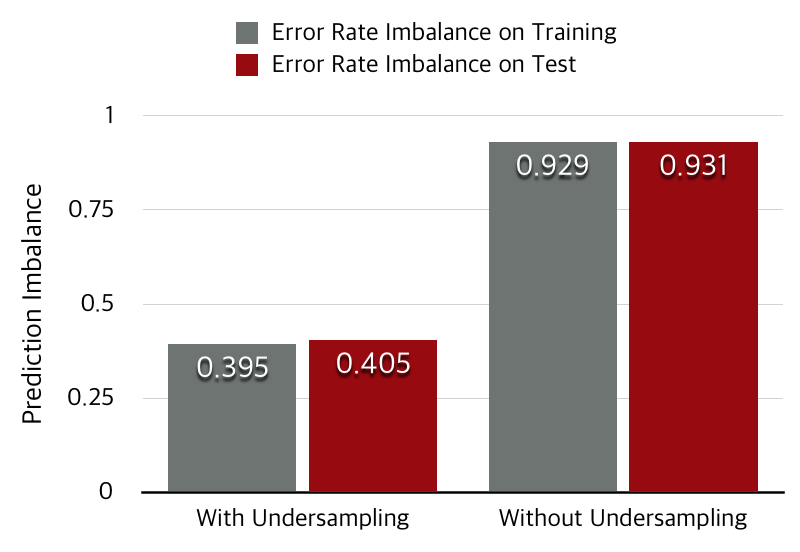}
    \caption{Prediction Imbalance between With Undersampling and Without Undersampling}
    \label{fig:imbalance_undersampling}
\end{figure} 

The results are as we expected, undersampling made both training and test AUC higher than not using undersampling, shown in Fig \ref{fig:auc_undersampling}. The strong need for using undersampling is shown in Fig \ref{fig:imbalance_undersampling}. We defined $$\text{Prediction Imbalance} =|err_+ - err_-|$$ to represent the magnitude of prediction imbalance where $err_+$ means error rate on predicting positive edges and $err_-$ on negative edges. If we let negative examples dominates positive examples, then the model will be overfitted to predict negative examples which results in serious bias of the model.

So we decided to use undersampling technique throughout our project.

\subsubsection{Area Under Curve (AUC)}

In order to evaluate the performance of our link prediction algorithm we will use one of the frequently used accuracy metrics which is known as Area Under Curve(AUC) statistics. AUC is calculated as the area under the Receiver Operational Curve (ROC).

In this study, the y axis of ROC curve will show the ratio of correctly identified true-positive types.  On the other hand, the x axis of ROC will show us the ratio of correctly identified true-negative types.  The range of y axis will be [0-1] since it is a ratio. Similarly, the range of x axis is also [0-1].  So, the area of a curve in this plot will also be in the range of [0-1].  Therefore, AUC will have a range of [0-1].
 
The values of AUC which are above 0.5 will indicate how the algorithm performs better than chance.  In the ideal case, the algorithm is expected to produce output where AUC = 1.

\section{Experiments And Analysis}
\subsection{Basic Analysis of Network}
We employed basically a number of SNA and machine learning tools, i.e. \textit{Snap.py, NetworkX, Gephi, IBM i2, $H_2O$ etc}.

\begin{figure}
\centering
        \includegraphics[totalheight=6cm]{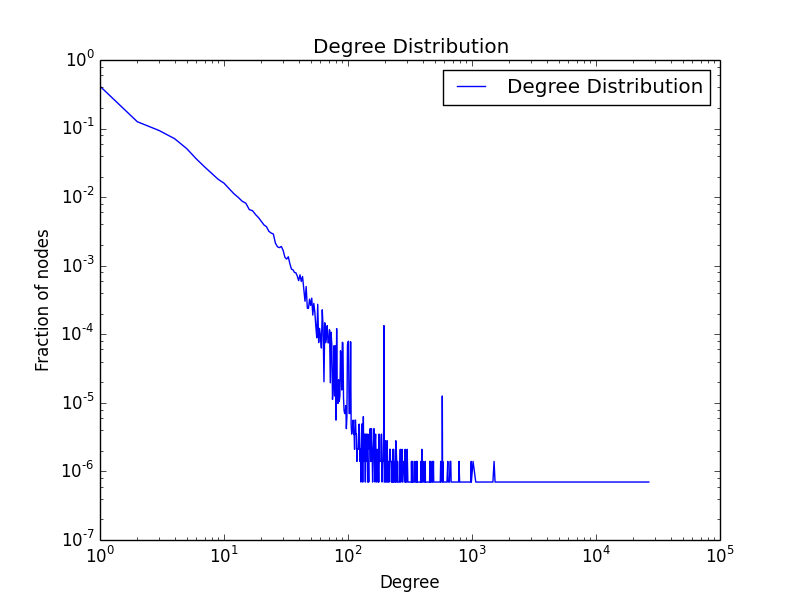}
    \caption{Degree Distribution}
    \label{fig:degreedistribution}
\end{figure}

\begin{figure}
\centering
        \includegraphics[totalheight=6cm]{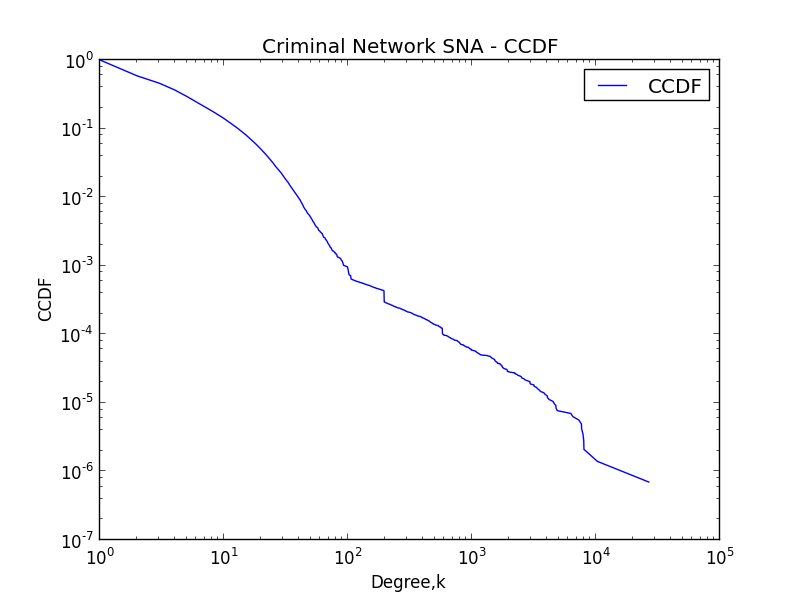}
    \caption{Complementary Cumulative Distribution Function(CCDF)}
    \label{fig:ccdf}
\end{figure}

As an initial work, we extracted degree distribution and complementary cumulative distribution function(CCDF) plot as shown in Figure \ref{fig:degreedistribution} and Figure \ref{fig:ccdf} respectively.  It is clearly seen on the plot that degree distribution follows power law with heavy tail structure.  We figured out the parameters of the power law fitting in our dataset as 
$\alpha=3.6345887074$ and $x_{min} =29.0$ which tell us that the criminal network follows Preferential Attachment Model.

In addition, we extracted a number of fundamental figures about the network properties as shown in the following table.

\begin{table}[h]
\centering
\begin{tabular}{@{}ll@{}}
\toprule
Characteristics                        & Value \\ \midrule
Nodes              & 1428002           \\
Edges                  & 3811872           \\
Bidirectional Edges & 7623744 \\
Closed triangles     & 18235211           \\
Open triangles                & 897960146           \\
Fraction of closed triads            & 0.019903           \\
Maximum WCC Size & 0.538763 \\
Maximum SCC Size & 0.538763 \\
Approximate full diameter & 29 \\
$90\%$ effective diameter & 11.736947 \\
Number of WCCs in the network &  160840 \\
\bottomrule
\end{tabular}
\vspace{0.2cm}
\caption{Characteristics of a criminal network}
\label{table:networkcharacteristic}
\end{table}

\subsection{Analysis on Link Prediction Model}
\subsubsection{How to choose negative examples for undersampling}

The set of negative examples in the criminal network is too large that we may not guarantee that choosing $|E^{+}|$ number of negative edges from the set $E^{-}$ which can be representative for the whole $E^{-}$.

So, we designed an experiment such that comparing randomly downsampled negative edges from $E^{-}$ with random sampled negatives from $E^{-}_d$ where $2\le d \le 10$. $E^{-}_d$ denotes the set of negative edges where the pair's shortest path length is $d$.

\begin{figure}
\centering
        \includegraphics[totalheight=6cm]{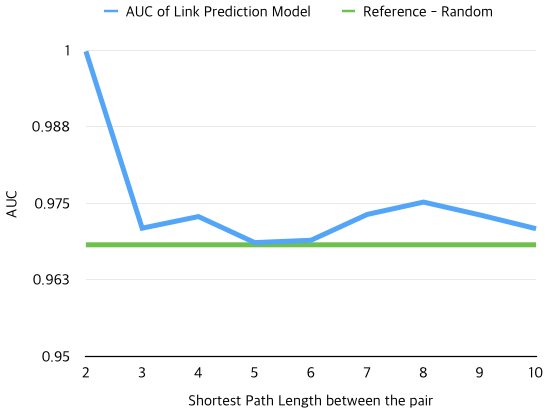}
    \caption{Performance of Link Prediction Model vs. Sampling of Negative Examples according to Pair's Shortest Path Length}
    \label{fig:random_sampling}
\end{figure}

The result is shown in Fig \ref{fig:random_sampling}. Interestingly, we can make a really accurate classifier if we use negative examples from $E^-_2$. Since the green reference line is the lower bound of the blue line, we could conclude that randomly downsampling of $E^-$ guarantees the minimum performance of the link prediction model and not overestimate the performance. So, we decided to downsample randomly from $E^-$ for our standard data sampling method for future experiments.

\subsubsection{Building Prediction Model and Feature Selection}

For the training phase of the classification task, we extracted some relevant features of our entities.  While reviewing some relevant features that are commonly used in the literature, we realized that Hasan et. al. proposed a variety of good features such as \textit{Common Neighbors, Jaccard Coefficient, Adamic/Adar, Katz} and so on  \cite{Hasan2011}. 

\begin{figure}
\centering
        \includegraphics[totalheight=5cm]{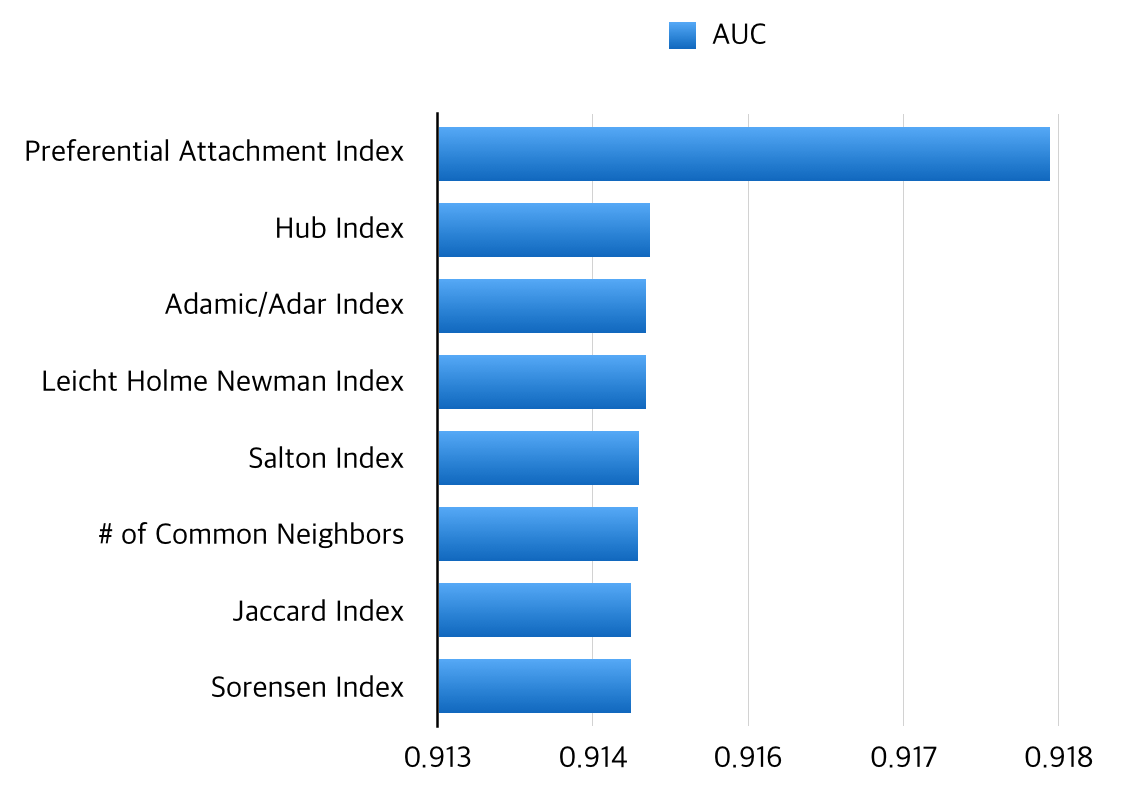}
    \caption{AUC when used one feature}
    \label{fig:auc_one_feature}
\end{figure}

We trained our model by computing various features on the dataset $D_{\text{train}}$. We tested our model on $D_{\text{test}}$ and obtained an AUC value.   We obtained an AUC value for each feature we applied. These AUC values can be seen in in Figure \ref{fig:auc_one_feature}.  The results show that discriminating power of preferential attachment index outweighed others. So we decided to choose \textit{preferential attachment index} as first feature for possible combination of features since it has the highest AUC value. 

\begin{figure}
\centering
        \includegraphics[totalheight=5cm]{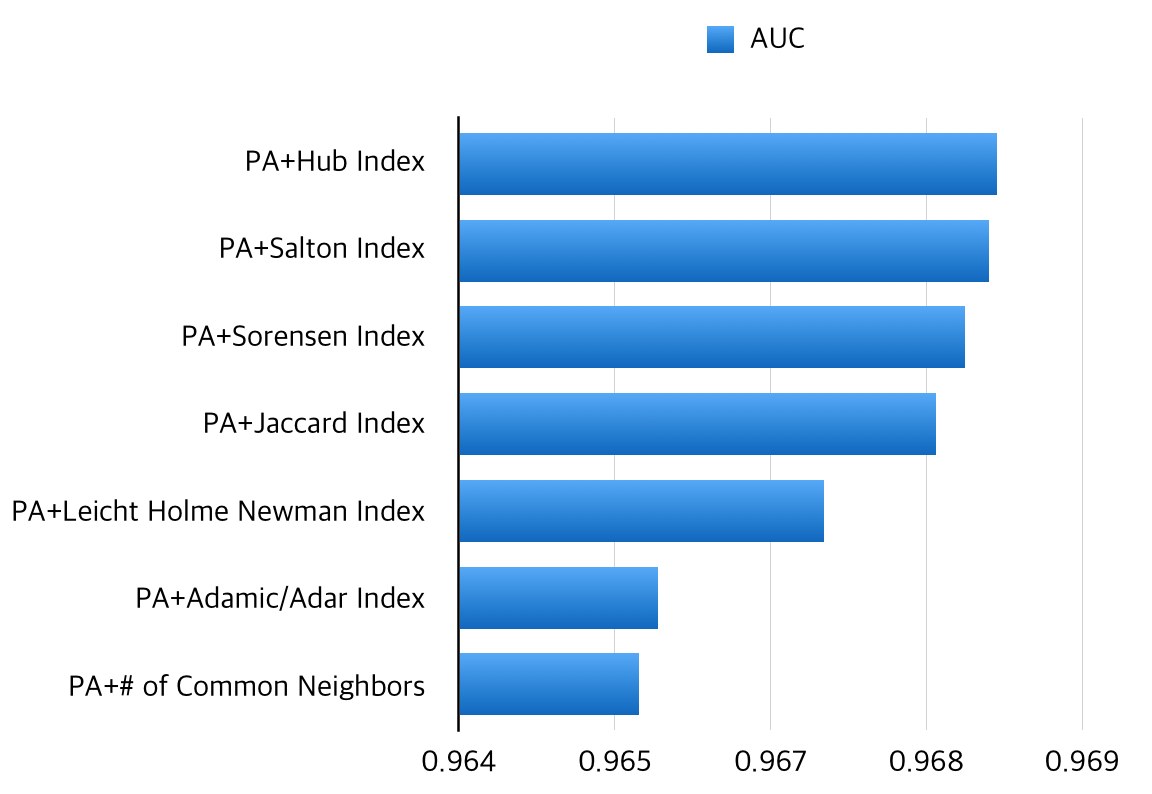}
    \caption{AUC when used two feature}
    \label{fig:auc_two_feature}
\end{figure}

Furthermore, we tested all features again as the second feature with preferential attachment index evaluated by AUC.  As a result, we obtained the bar chart shown in Figure \ref{fig:auc_two_feature}. It is clearly seen that the inclusion of hub index as a second feature improves AUC.  We tried the inclusion of each feature as the third feature as well, but from adding the third feature, we observed little improvement by inclusion of additional features on AUC result. So, preferential attachment index and hub index were the most important features and we trained and finalized our prediction model with AUC of 0.96822.

\subsubsection{Probability of Hidden Link vs. Shortest Path Lengths}
In order to find out the relationship between the probability of possible hidden link and shortest path length between two nodes, we had to randomly sample 9 sets of $E^{-}_d$ where $2\leq d \leq 10$. In other words, we made 9 different datasets of negative edges clustered by shortest path length of range from 2 to 10. These data sets were independently drawn from $E^-$ to the negative edges used in $D$.

We expected a large probability of a hidden link when the pair has small $d$, due to triadic properties. Naturally, we expected low probability of a hidden link when the pair has large $d$. The probabilities can be calculated by applying our prediction model on all datasets denoted previously as $E^{-}_d$.
	

\begin{figure}
\centering
        \includegraphics[totalheight=5cm]{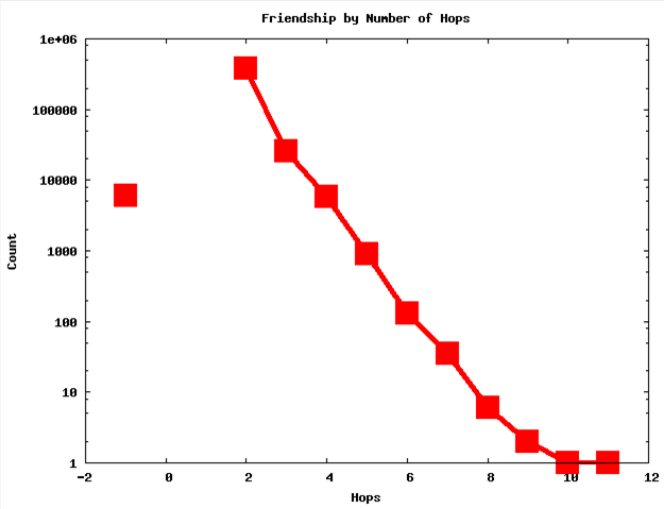}
    \caption{Triad Closure Behaviour of Facebook Network:  Distance x=2 denotes the friends of
friends while x=-1 denotes isolated  nodes\cite{Backstrom_2011}}
    \label{fig:jure_triad_closure}
\end{figure}

\begin{figure}
\centering
        \includegraphics[totalheight=5cm]{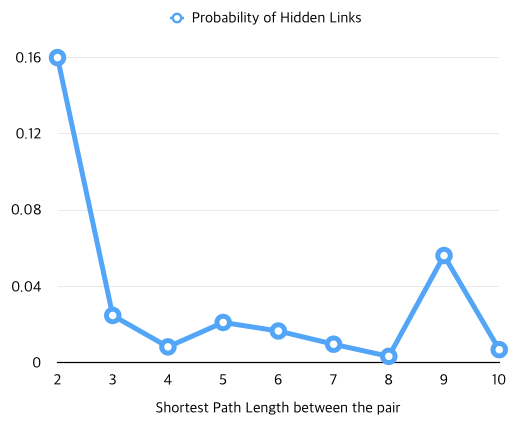}
    \caption{Probability of Hidden Links vs Shortest Path Length}
    \label{fig:error_vs_shortest_path}
\end{figure}
\begin{figure}
\centering
        \includegraphics[width=7cm]{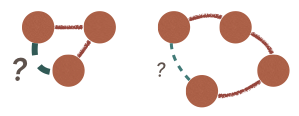}
    \caption{Concept diagram for Probability of Hidden Links vs. Shortest Path Length}
    \label{fig:diagram_spl}
\end{figure}

As we expected, Figure \ref{fig:error_vs_shortest_path} shows that the negative edges with the shortest path length of 2 have the highest probability of being a hidden link. The result is very well aligned with the triadic closure property of Facebook network revealed by Leskovec et.al. in Fig \ref{fig:jure_triad_closure} \cite{Backstrom_2011}. Generally speaking, the tendency is the shorter the shortest path length, the higher the probability of hidden links as shown in Fig \ref{fig:diagram_spl}.

\subsection{Finding Hidden Links on Corrupted Networks}

For preparing the corrupted networks, we randomly chose edges of portions $5\%, 10\%,...,50\%$  and remove them from the original network. Let's denote a particular corrupted network as $C_p$ and removed positive edges as $R_p$, where $p \in \{5\%,10\%,15\%...,50\%\}$ denotes portion of removal.

We designed two experiments. One is to find out how well machine learning models can be built in corrupted networks with respect to the portion of removal while the latter one is to test how well the built model can reconstruct the original network. We thought the latter is important since we have an incomplete criminal network and there might be lots of hidden links and our goal is to find those hidden links. So the performance of reconstruction may be the measure of how well we can reveal hidden links from incomplete information (network).

\subsubsection{Performance of Link Prediction Model from Corrupted Network}
From each corrupted network, $C_p$, we made a training and test data sets, namely $D_{\text{train},p}, D_{\text{test},p}$ where $p \in \{5\%,10\%...,50\%\}$ using similar ways to train the original network. So, we have distinct link prediction model for each corrupted network.

\begin{figure}
\centering
        \includegraphics[totalheight=5cm]{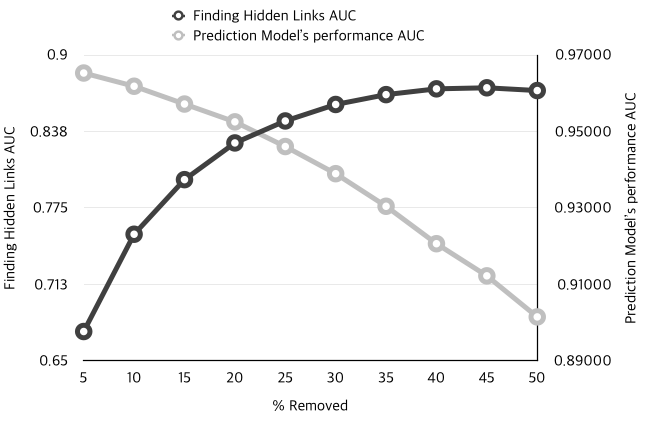}
    \caption{AUC vs \% Edges Removed}
    \label{fig:auc_vs_removed_edge_ratio}
\end{figure}

\begin{figure}
\centering
        \includegraphics[totalheight=5cm]{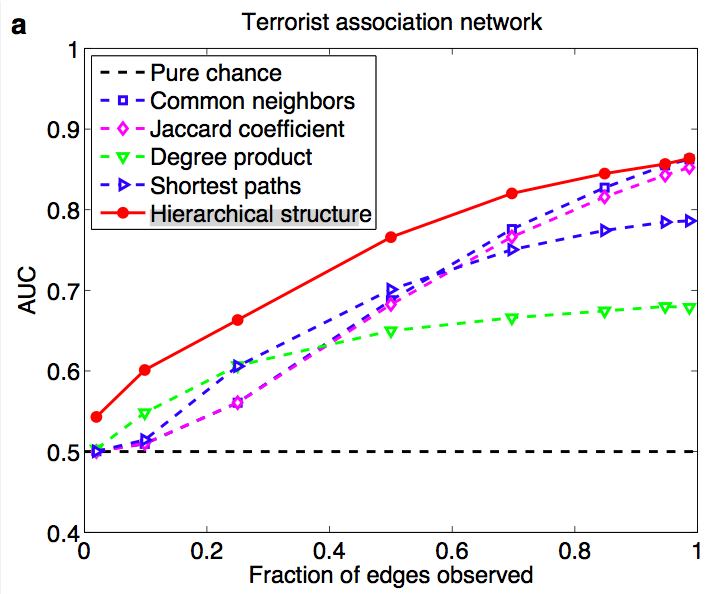}
    \caption{AUC vs Fraction of Edge Observed \cite{Clauset2008}}
    \label{fig:edge_removal_performance}
\end{figure}

For each corrupted network $C_p$, we obtained an AUC value on $D_{\text{test},p}$ using a model trained on $D_{\text{train},p}$. Gray line in Fig  \ref{fig:auc_vs_removed_edge_ratio} shows that AUC values of models in corrupted networks have decreasing tendency when $p$ gets larger. This result shows similarity to the result of past research which can be seen in Fig  \ref{fig:edge_removal_performance}. Clauset et. al. used just one link prediction metric to obtain the best AUC of their study\cite{Clauset2008}. Since we used multiple prediction metrics in our model, we achieved higher AUC than their results.

\subsubsection{Performance on Finding Hidden Links}
In this part of the experiment, we tested the performance of our model with a series of analysis as follows. The result is quite interesting. 

\begin{figure}
\centering
        \includegraphics[totalheight=5cm]{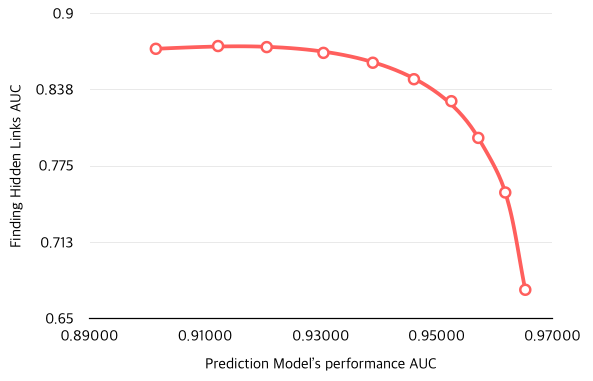}
    \caption{AUC on Finding Hidden Links vs. Prediction Model's AUC}
    \label{fig:train_vs_test}
\end{figure}

As we mentioned in the previous section, we tested how well each model from a corrupted network can be reconstructed the corresponding removed edges. The result was shown in the black line in Fig \ref{fig:auc_vs_removed_edge_ratio}. The resulting plot shows increasing accuracy in finding hidden links when more edges were removed. This result is the exact opposite to one of model's AUC in gray line in the same figure. The opposite trends can be clearly shown when we draw additional graph in Fig \ref{fig:train_vs_test}.

Since removal of edges make the graph more incomplete, we expected decrease in AUC on finding hidden links when the ratio of removed edge increase. However, the observed behavior is surprisingly different from our expectation.  

\begin{figure}
\centering
        \includegraphics[totalheight=5cm]{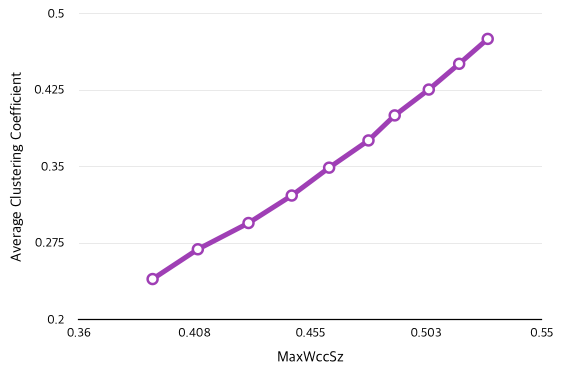}
    \caption{Average Clustering Coefficient vs. Maximum WCC Size}
    \label{fig:avgcf_vs_maxwccsz}
\end{figure}

We guessed there must be more clues for this result such as the network structures must be related to the performance of finding hidden links. So, we investigated three representative measurements of the network: Sampled Diameter, Maximum WCC Size, and Average Clustering Coefficient.
\begin{figure}
\centering
        \includegraphics[totalheight=5cm]{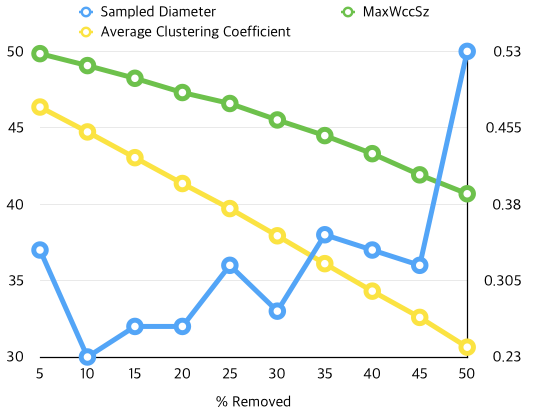}
    \caption{Various Network Measurements vs. \% Edges Removed}
    \label{fig:statistics}
\end{figure}
Fig \ref{fig:statistics} shows measurements of the network with respect to the portion of edges removed in percentage. 
Sampled diameter fluctuates, but its moving average increases as more edges are removed. On the other hand, Maximum WCC Size and Average Clustering Coefficient  decreased steadily, as more edges are removed. It can be intuitively validated that the more edges are removed, the more sparse the network becomes, hence, leading to a decrease both in clustering coefficient and the size of the largest WCC. We can further explain this tendency by Fig \ref{fig:avgcf_vs_maxwccsz}, which shows that there is a linear  dependency between average clustering coefficient and maximum WCC size and removal of edges reduces both of them.

\begin{figure}
\centering
        \includegraphics[totalheight=5cm]{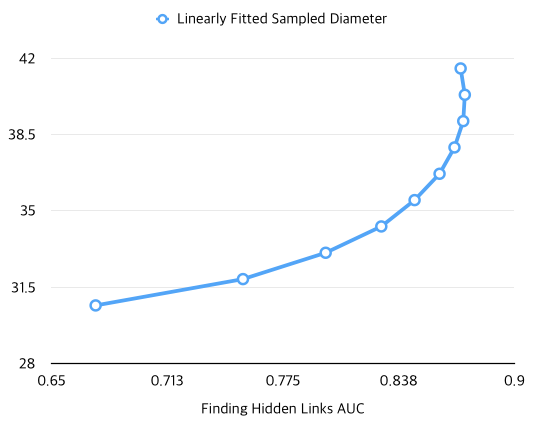}
    \caption{Linearly Fitted Sampled Diameter vs. Finding Hidden Links AUC}
    \label{fig:diameter_vs_auc}
\end{figure}

\begin{figure}
\centering
        \includegraphics[totalheight=5cm]{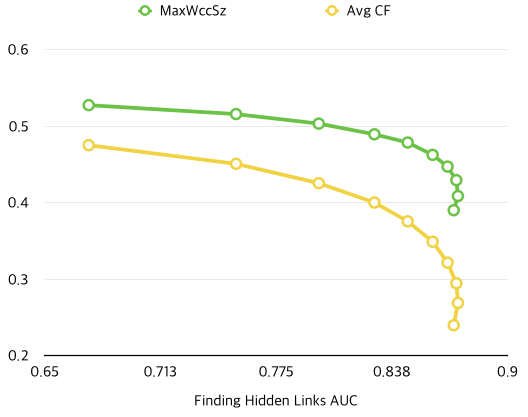}
    \caption{Maximum WCC Size and Average Clustering Coefficient vs. Finding Hidden Links AUC}
    \label{fig:maxwccsz_avgcf_vs_auc}
\end{figure}

\begin{figure}
\centering
        \includegraphics[width=7cm]{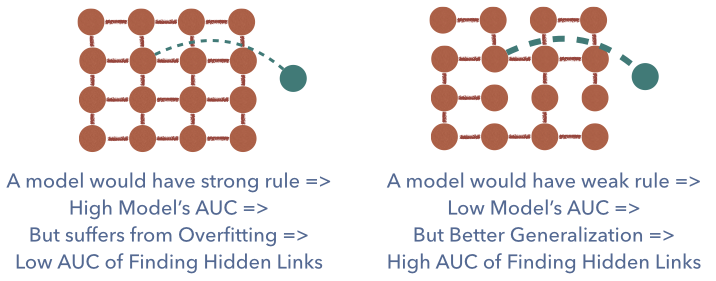}
    \caption{Concept diagram for Network Characteristics and Performance on Finding Hidden Links}
    \label{fig:concept_overfitting}
\end{figure}

And finally, the interesting result in Fig \ref{fig:train_vs_test} can now be further explained by characteristics of the network as shown in Fig \ref{fig:diameter_vs_auc} and \ref{fig:maxwccsz_avgcf_vs_auc}. Those three plots of performance on finding hidden links versus linearly fitted sampled diameter, average clustering coefficient, and maximum WCC size gives us consistent result and let us analyze with the characteristics of the network. Our criminal network has property that if maximum WCC size decreases, then the average clustering coefficient will also decrease. And since decreased maximum WCC size means more segmentation of WCC's throughout the network, diameter would increase as maximum WCC size decreases. Fig \ref{fig:diameter_vs_auc} and \ref{fig:maxwccsz_avgcf_vs_auc} shows that in this case, the AUC of finding hidden links increases while the AUC of prediction model decreases. We can explain this phenomenon using this with Fig \ref{fig:concept_overfitting}:

\begin{itemize}
\item If the network has a larger core and denser structure (large maximum WCC size, large average clustering coefficient, and small diameter), this will make the prediction model learn more about the current network(higher performance of prediction model).
\item However, in this case, since we can say the prediction model has been overfitted to the current image of network, the performance of finding hidden links(which are not real edges on corrupted networks) would decrease.
\end{itemize}
As a result of this series of analysis, it is observed that pruning out some fraction of edges from the original network would help achieve better performance in finding hidden links.

\subsection{Destroying Criminal Networks}

One of our final goals is to find out a smart way of destroying criminal networks. So, we focused on what the most effective way is for arresting criminals to destroy the network in shorter time. The intuitive way to do that is to choose random node and remove it. The better way is to choose node with highest degree first. More sophisticated way is to choose node with highest pagerank score first. 

At this point, we will use our prediction model to give weights on existing edges by probabilities of being a link between two nodes. More suspicious connections will have higher weights(close to 1) and less suspicious ones will have lower weights(close to 0). Using those weights we could create weighted network and get the weighted pagerank scores.

We will do the robustness test using those weighted pagerank scores. We expect a faster collapse of the criminal network using our method.

\subsubsection{Performance of Weighted PageRank method in Destroying the Network}

\begin{figure}
\centering
        \includegraphics[totalheight=5cm]{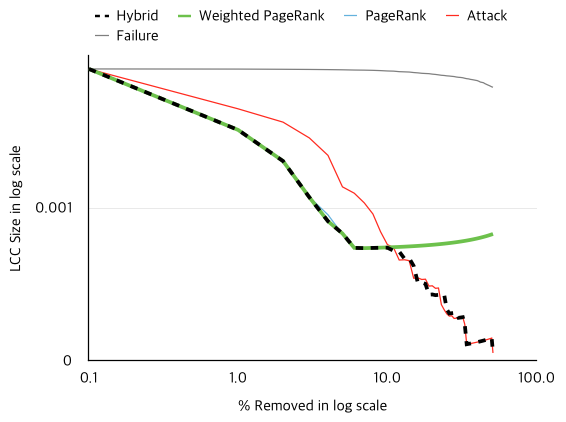}
\includegraphics[totalheight=5cm]{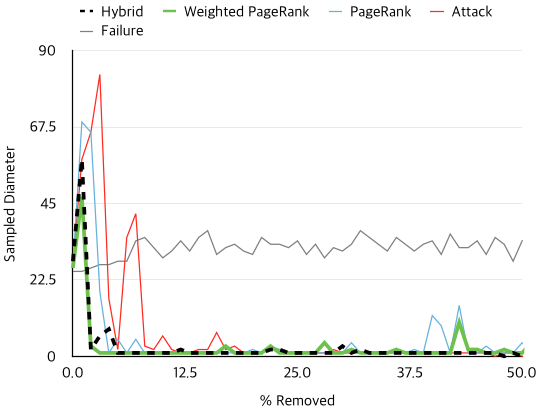}
    \caption{Destruction Characteristics of Criminal Network in Different Attack Scenarios}
    \label{fig:attack_scenario}
\end{figure}

We calculated the classical PageRank scores of each node based on unweighted edges on the original network.  In addition, we calculated \textit{Weighted PageRank} scores based on the weights we obtained.  We analysed the change of diameter and the ratio of largest connected component under a random failure scenario and 3 targeted attack scenarios based on degree of nodes, PageRank, and Weighted PageRank scores.  The results are shown in
Figure \ref{fig:attack_scenario}.

We observed in the first plot of Figure \ref{fig:attack_scenario} that removal of nodes having maximum Weighted PageRank score reduces the largest component of the network much faster than removal of nodes having maximum degree.  It means the Weighted PageRank scores we obtained identifies the most influential nodes of a network much precisely.  So, in the criminal network context, we called the Weighted PageRank score as \textit{suspiciousness index}. 
Consequently, suspiciousness index can be used to take a pareto approach to maximize preventive risk reduction of possible criminal activities by watching the \textit{vital few} of the network much more closely.

On the other hand we observed no clear difference between effects of PageRank and Weighted PageRank scores on the destruction characteristics of criminal network in the LCC size plot of Fig. \ref{fig:attack_scenario}.  However, we observed clear difference on the influence of PageRank and Weighted PageRank in the diameter plot of Fig. \ref{fig:attack_scenario}. We also validated our expectation by observing attack scenarios outweighed the random failure scenario.

\subsubsection{Comparing Weighted PageRank Method to Single Index PageRank Methods}
Using our prediction model, we can obtain probabilities of each edge being a link.  The probabilities are the output of our machine learning algorithm which uses an ensemble of indexes.  We use these probabilities to turn our original network into weighted network.  Then, we used the weights as an input to the pagerank algorithm, which we called \textit{weighted pagerank method}, to obtain \textit{weighted pagerank scores} of the nodes. 

\begin{figure}
\centering
        \includegraphics[totalheight=5cm]{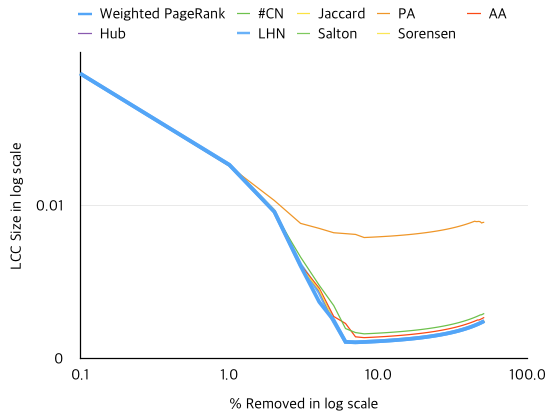}
\includegraphics[totalheight=5cm]{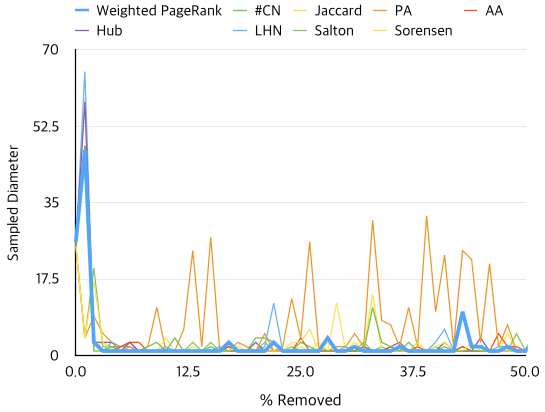}
    \caption{Comparing Weighed PageRank Method to Single Index PageRank Methods}
    \label{fig:single_index}
\end{figure}

We came up with the idea that since our prediction model uses ensemble of multiple indices, it must outperform any single index pagerank methods which uses single index, i.e. Jaccard index.  As it can be seen in Fig \ref{fig:single_index} with a blue bold line, the weighted pagerank method outweighs other pagerank methods which uses single index. So, we achieved the maximum performance in destroying criminal network by using the weighted pagerank method.

\subsubsection{Hybrid Method for the Best Performance in Destroying the Network}
As it can be seen in Fig \ref{fig:attack_scenario}, all pagerank methods have a performance bottleneck after some period of attacks. This may be because we calculate the pagerank score only at the very first time and the subsequent snapshots of attacked network were not reflected in the pagerank scores. If we can calculate pagerank score at every iteration step, it would yield different results and may even be the maximum performance. However, since calculating pagerank scores for each node in every snapshot is very expensive, we were not able to observe that result.  As another practial solution, we designed a new method which can combine the advantages of both weighted pagerank method and one that chooses a node with highest degree first at each round. The result of this hybrid method is shown in Fig. \ref{fig:attack_scenario} with black dashed line. Its performance in terms of maximum WCC size exactly meets our expectation.
$$W_{\text{hybrid}}\approx \min{(W_{\text{Weighted Pagerank}}, W_{\text{Attack}})} $$
Also, in terms of sampled diameter, hybrid method has heavy tail and fast decreasing tendency at the first attack phase which means it has the advantages of both pagerank and highest degree first method, hence, achieves the maximum performance.

As a result, we achieved the maximum performance on destroying the criminal network, by using the hybrid method.

\section{Conclusion}
\subsection{Link Prediction Model}
We applied GBM to make a machine learning model which can describe the current network. We used indices which are famous in link prediction algorithms as features of the model. Instead of employing single index, we ensembled multiple indices to improve prediction performance.

\subsection{Finding Hidden Links}
We tested the performance of our prediction model to find hidden links in the corrupted networks. From our experiments, we concluded that the more decentralized and less clustered network has better performance on finding hidden links. This is because our machine learning model avoids overfitting to our dataset and has low variance as the dataset becomes less clustered and more decentralized by removing more edges. 


\subsection{Destroying the Network}
Finally, we used the probabilities predicted by the model on current edges as weights of the network, and obtained weighted pagerank scores from the weighted network. We tested various attacking methods and the two methods, weighted pagerank method and removing the one with highest degree method, had both advantages and disadvantages. At the end we combined advantages of both methods and achieved the best performance on destroying network.

\section{Future Research}
\subsection{Applying various Machine Learning Techniques}
We only used GBM as our machine learning technique. However, there are many other potential techniques we could apply. We can apply logistic regression, random forest, SVM, and neural network to get the various results.
Also, GBM has parameters to be tuned. So we could do a grid search to find the best parameters for link prediction model.

\subsection{Finding additional indices to be used for features}
One of the most important issues in machine learning problem is finding useful features. We used various indices to our model, but there might be useful additional features such as clustering coefficients, pagerank scores, and triadic ratio.
We may apply those indices to improve link prediction model.

\subsection{Comparison of our Link Prediction Model to Existing Methods}
We could not do an experiment on temporal change of our data set since there was only a  snapshot of the criminal network data set at the time of the study. If we could obtain the temporal changes of our data set, we would be able to compare the performance of our predictions to real data.  In addition, we would be able to compare our prediction performance with the prediction performance of the study carried by Leskovec et.al.\cite{Backstrom_2011}.

\subsection{Using Actual Weights of the Network}
The network only has information whether there is an edge between the pair of nodes or not. If we have additional information such as the amount of phone calls, transactions, and number of suspicious meetings, we can use them as additional weights for the edges in the network. Using this weighted network, we will be able to use also regression models instead of classification model. Also, we can compare the network destruction performance of our weighted pagerank method by using the new weights and existing weights, respectively. 

\bibliography{bibliography}

\begin{thebibliography}{10}

\bibitem{Backstrom_2011}
Lars Backstrom and Jure Leskovec.
\newblock Supervised random walks: Predicting and recommending links in social
  networks.
\newblock In {\em Proceedings of the Fourth ACM International Conference on Web
  Search and Data Mining}, WSDM '11, pages 635--644, New York, NY, USA, 2011.
  ACM.

\bibitem{Backstrom:2011:SRW:1935826.1935914}
Lars Backstrom and Jure Leskovec.
\newblock Supervised random walks: Predicting and recommending links in social
  networks.
\newblock In {\em Proceedings of the Fourth ACM International Conference on Web
  Search and Data Mining}, WSDM '11, pages 635--644, New York, NY, USA, 2011.
  ACM.

\bibitem{Clauset2008}
Aaron Clauset, Cristopher Moore, and M.~E.~J. Newman.
\newblock Hierarchical structure and the prediction of missing links in
  networks.
\newblock {\em Nature}, 453:98--101, 2008.

\bibitem{farnaz2003}
Farnaz Fassihi.
\newblock Two novice gumshoes charted the capture of {S}addam {H}ussein,
  December 2003.

\bibitem{gbmbook}
Trevor Hastie Robert~Tibshirani Gareth~James, Daniela~Witten.
\newblock {\em An Introduction to Statistical Learning with Applications in R}.
\newblock Springer.

\bibitem{icpvtr2014}
Rohan Gunaratna.
\newblock Icpvtr - international centre for political violence and terrorism
  research, 2014.

\bibitem{Hasan06linkprediction}
Mohammad~Al Hasan, Vineet Chaoji, Saeed Salem, and Mohammed Zaki.
\newblock Link prediction using supervised learning.
\newblock In {\em In Proc. of SDM 06 workshop on Link Analysis,
  Counterterrorism and Security}, 2006.

\bibitem{Hasan2011}
Mohammad~Al Hasan and MohammedJ. Zaki.
\newblock A survey of link prediction in social networks.
\newblock In Charu~C. Aggarwal, editor, {\em Social Network Data Analytics},
  pages 243--275. Springer US, 2011.

\bibitem{Liben2003}
David Liben-Nowell and Jon Kleinberg.
\newblock The link prediction problem for social networks.
\newblock In {\em Proceedings of the Twelfth International Conference on
  Information and Knowledge Management}, CIKM '03, pages 556--559, New York,
  NY, USA, 2003. ACM.

\bibitem{Medina2014}
R.~M. Medina.
\newblock Social network analysis: A case study of the islamist terrorist
  network.
\newblock {\em Secur J}, 27(1):97--121, Feb 2014.

\bibitem{ofac}
Office of~Foreign Assets~Control.
\newblock Specially designated nationals list (sdn), October 2014.

\bibitem{Sparrow1991}
Malcolm~K. Sparrow.
\newblock The application of network analysis to criminal intelligence: An
  assessment of the prospects.
\newblock {\em Social Networks}, 13(3):251 -- 274, 1991.

\bibitem{decision_tree_wiki_2004}
Wikipedia.
\newblock Decision tree learning, 2014.

\bibitem{wiki:xxx}
Wikipedia.
\newblock Oversampling and undersampling in data analysis --- wikipedia{,} the
  free encyclopedia, 2014.
\newblock [Online; accessed 5-December-2014].

\end{thebibliography}
\bibliographystyle{plain}
\nocite{*}

\end{document}